# Structure and superconductivity of Mg(B$_{1-x}$C$_x$)$_2$ compounds

Shao-ying Zhang, Jian Zhang, Tong-yun Zhao, Chuan-bing Rong, Bao-gen Shen and Zhao-hua Cheng*

State Key Laboratory of Magnetism, Institute of Physics and Center of Condensed Matter Physics, Chinese Academy of Sciences, Beijing 100080, P.R. China

**Abstract:**

In this paper, we reported the structural properties and superconductivity of Mg(B$_{1-x}$C$_x$)$_2$ compounds. Powder x-ray diffraction results indicate that the samples crystallize in a hexagonal AlB$_2$-type structure. Due to the chemical activity of Mg powders, a small amount of MgO impurity phase was detected by x-ray diffraction. The lattice parameters decrease slightly with increasing carbon content. Magnetization measurements indicate the non-stoichiometry of MgB$_2$ has no influence on the superconducting transition temperature and the transition temperature width. The addition of carbon results in a decrease of T$_c$ and an increase in the superconducting transition width, implying the loss of superconductivity.

*Corresponding author

## I. Introduction

The discovery of superconductivity at about 39 K in magnesium diboride by Akimitsu et al. has resulted in a common excitement in the scientists investigating in the field of solid state physics due to its relatively high superconducting transition temperature[1]. Its superconducting transition temperature seems to be at or above the limit of BCS theoretical model[2]. This novel discovery may open a new route towards the search of high temperature superconductors. In order to investigate the effect of electron concentration on $T_c$, Slusky et al. have synthesized the solid solutions of $Mg_{1-x}Al_xB_2$ and investigated its structural transition and superconductivity[3]. It was found that the superconducting transition temperature decreases with increasing Al concentrations, and the superconductivity disappears with x>0.4. Up to data, the contribution of B layers to the superconductivity is not yet well known. The substitution of carbon for boron may give insight into understanding the contribution of boron layer to the superconductivity. In this letter, the effect of carbon substitution on the superconductivity of $MgB_2$ was investigated.

## II. Experimental

The samples were prepared from powdered magnesium (98.5% in purity), B(99.99% in purity) and C(99.99% in purity) powers. The powders were well mixed in an appropriate ratio, and pressed into pellets. The pellets were sealed in a quartz tube and sintered at 950 K for two hours. Powder x-ray diffraction was carried out by a x-ray spectrometer with Cu $K\alpha$ radiation over a 2θ from 20° to 90° with a step of

0.02°. The temperature dependence of magnetization of the samples was measured in a Quantum Design MPMS-7 SQUID magnetometer in an applied field of 20 Oe.

**III. Results and Discussion**

Fig.1 illustrates the x-ray diffraction patterns of $MgB_2$ and $Mg(B_{1-x}C_x)_2$. The XRD patterns indicate that the main diffraction peaks can be indexed with a hexagonal structure of $Mg(B_{1-x}C_x)_2$. A small amount of impurity phase MgO was detected due to the chemical activity of Mg powders. The crystal structure of $MgB_{1.8}C_{0.2}$ consists of alternating hexagonal layers of Mg atoms and graphite-like honeycomb layers of B and C atoms along c-axis direction. With introduction of carbon, the lattice parameters (a=3.070 A, c=3.523A) are slightly smaller than those of $MgB_2$ ($a$=3.086A, $c$=3.524A).

Fig.2 illustrates the temperature dependence of magnetization of $Mg(B_{1-x}C_x)_2$ intermetallic compounds measured in applied field of 20 Oe. A sharp superconducting transition was observed in $MgB_2$ compound. With increasing carbon content, the temperature dependence curves become less sharp. The superconducting transition temperature determined from the onset temperature of 2% of the full diamagnetic signal at low temperature was found to decrease from 36 K for x=0.00 to 34K for x=0.20. The superconducting transition width for a 10 to 90% drop, ΔTc increases from 1 K to 3 K with introducing carbon atoms.

The superconducting transition temperature of our sample $MgB_2$ is slightly lower than 39 K, as reported in refs 1 and 3. One possible reason for lower Tc is related to the non-stoichiometry of $MgB_2$. It is well known that boron and carbon have high melting points and high boiling points, the only possible loss is Mg powders due to its evaporation at high temperatures. In order to exclude the possibility of deficiency of Mg, we have investigated the non-stochiometry of $MgB_2$ by adding an excess of Mg with 10at.%, 20at.% and 30at.% Mg, respectively. Figure 3 presents the temperature of magnetization of $Mg_{1+x}B_2$ (x=0.00, 0.1,0.2 and 0.3) compounds. In order to compare each other easily, the magnetization data are all normalized to 1 at low temperatures. The plot shows that they have identical superconducting transition temperature and transition width regardless of Mg contents. Therefore, the lower value of Tc for our samples is possibly resulted from the low impurity of Mg powders(98.5% in purity)., rather than the non-stoichiometry of $MgB_2$ compound.

## IV. Conclusion

The structural properties and superconductivity of $Mg(B_{1-x}C_x)_2$ were investigated by means of powder x-ray diffraction patterns and magnetization measurement. It was found that the non-stoichiometry of $MgB_2$ has no influence on supercondcuting transition temperature and its width. This result suggests that once $MgB_2$ forms, it has specific Tc. The deficiency or excess of Mg only results in a formation of second phase, which has no effect on Tc. The addition of carbon leads to decrease the supeconducting transition temperature.

**Acknowledgements:**

This work was supported by the National natural Sciences foundation of China and the State Key Project of fundamental research, China. The authors would like to thank Mr. Rui-wei Li and Mr. Tai-shan Ning for their kind help in sample preparation and x-ray diffraction measurements.

**References**

[1] J. Akimitsu, Symposium on Transition Metal Oxides, Sendai, Janary, 10, 2001; J. Nagamatsu, N. Nakagawa, T. Muranaka, Y. Zenitani and J. Akimitsu, Nature, **410**, 63(2001).

[2] W. L. McMillan, Phys. Rev. **167**,331(1968).

[3] J.S. Slusky, N. Rogado, K.A. Regan, M.A. Hayward, P.Khalifah, T. He, K Inumaru, S. Loureiro, M.K. Haas, H.W. Zanbergen and R.J. Cava, Loss of superconductivity and structural transition in $Mg_{1-x}Al_xB_2$, Nature (submitted)

**Figure captions:**

Fig. 2. Normalized magnetization data of $Mg(B_{1-x}C_x)_2$ compounds with x=0.00, 0.10 and 0.20 as a function of temperature measured in an applied field of 20 Oe. The samples were cooled down to 5 K without magnetic field(ZFC).

Fig. 3. The temperature dependence of normalized magnetization of $Mg_{1+x}B_2$ compounds measured in an applied field of 20 Oe. The samples were cooled down to 5 K in zero magnetic field.

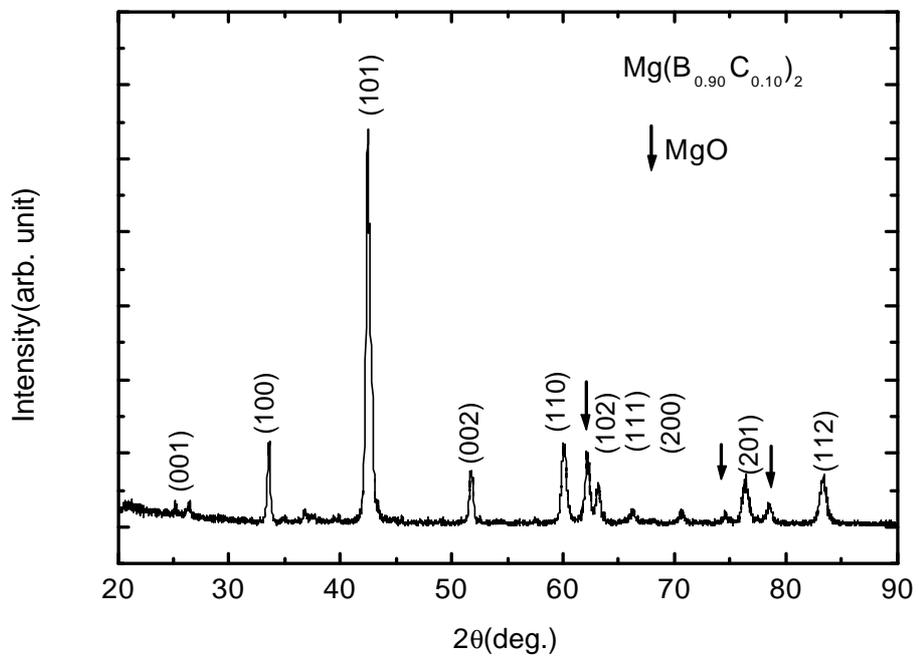

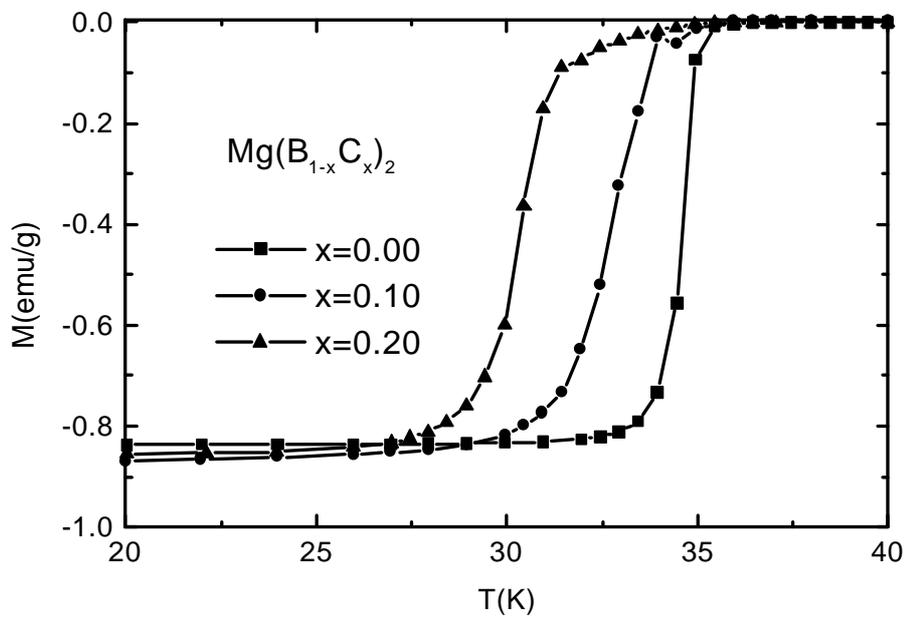

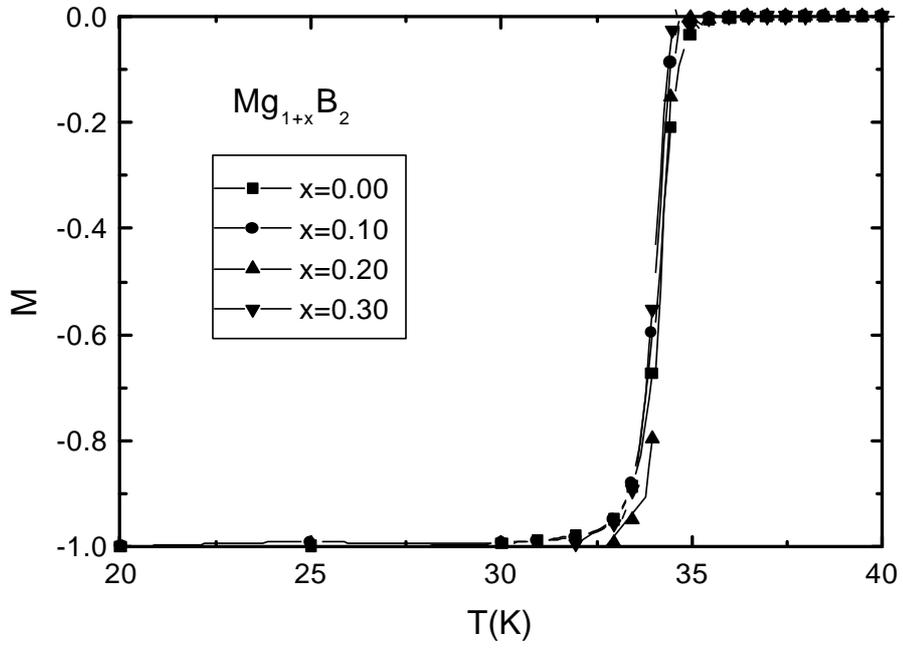